\documentclass[aps,pra,twocolumn,floatfix,showpacs,groupedaddress]{revtex4}
\usepackage{graphicx}

\usepackage{amsbsy,amssymb,amsmath,bm}

\begin{document}
\title{Stability of low-dimensional multicomponent dilute Bose gases}
\author{Alexei K. Kolezhuk}
\thanks{On leave from: Institute of
Magnetism, National Academy of Sciences and Ministry of Education,
03142 Kiev, Ukraine.}
\affiliation{Institut f\"ur Theoretische Physik C, RWTH Aachen University, 52056 Aachen,
  Germany}
\affiliation{JARA J\"ulich-Aachen Research Alliance, 
Research Centre J\"ulich GmbH, 52425 J\"ulich, Germany}

\begin{abstract}
I show that  in low dimensions the interactions in dilute Bose
mixtures are strongly renormalized, which
leads to a considerable change of stability
conditions  compared to the mean-field results valid in the high-density
regime. Estimates are given for the two-component Bose-Hubbard model and for the
$\rm{}^{87} Rb$-$\rm{}^{41}K$ mixture.
\end{abstract}
\pacs{03.75.Mn,67.85.Fg,67.60.Bc,37.10.Jk}

\maketitle

\section{Introduction}

Soon after achieving the Bose-Einstein condensation in ultracold dilute
atomic gases \cite{AnglinKetterle02rev}, there came a surge of interest in
degenerate quantum gas mixtures that has been continuing unabated up to the
present days. Multicomponent Bose condensates were initially
realized \cite{Myatt+97,Hall+98} by using two different hyperfine states of $\rm
{}^{87}Rb$, later miscible and immiscible spinor condensates of $\rm {}^{23}Na$
were observed \cite{Stamper-Kurn+98}. Heteronuclear condensates of $\rm
{}^{41}K$ and $\rm {}^{87}Rb$ have been realized in a trap \cite{Modugno+02} and
in an optical lattice \cite{Catani+08}. Very recently, different
groups have used Feshbach resonances to engineer two-component Bose condensates
with tunable inter- and intra-species interactions in
$\rm{}^{87}Rb$-$\rm{}^{39}K$ \cite{Roati+07}, $\rm{}^{87}Rb$-$\rm{}^{41}K$
\cite{Thalhammer+08}, and $\rm{}^{87}Rb$-$\rm{}^{85}Rb$ \cite{Papp+08} mixtures.

The interest to multicomponent Bose systems is driven, particularly, by their rich
physics: they provide an opportunity to study various phase transformations
ranging from demixing of interpenetrating quantum liquids
\cite{HoShenoy96,Timmermans98} to transitions between different magnetic states
in spinor condensates \cite{Ho98,OhmiMachida98}, and to a variety of yet
unobserved quantum phase transitions predicted to appear in the presence of an
optical lattice
\cite{Altman+03,Kuklov+04,Isacsson+05,ZhouSnoek03,ImambekovLukinDemler03}.  Both
in traps and optical lattices, the realization of highly anisotropic geometries
(``cigars'' and ``pancakes'') is possible \cite{Bloch+rev08}, which provides an
opportunity to study the above phenomena in lower-dimensional systems.

The simplest phase transition in a Bose mixture is the phase separation
(demixing) in a two-component system, which occurs if the inter-species
repulsion overcomes the intra-species one.  Existing theoretical studies of this
instability are usually performed at the mean-field level by using coupled
Gross-Pitaevskii equations \cite{HoShenoy96,Timmermans98}.  
It is well known
that in lower dimensions  mean-field arguments might become inapplicable in
the low density regime  \cite{PitaevskiiStringari-book}. In a one-dimensional Bose gas, with the
decrease of the density, the healing
length $\xi$ becomes smaller than the average interparticle distance $d$, thus
invalidating the mean-field approach. In two dimensions, the situation
is more subtle: although the ratio $\xi/d$ does not depend on the density and
falls below unity only in the case of very tight two-dimensional
confinement, in the dilute limit the effective coupling constant
becomes strongly energy- and density-dependent \cite{PetrovHolzmannShlyapnikov00,Lee+02}.

For
one-dimensional (1D) multicomponent systems, a study beyond the mean-field approximation, based
on the bosonization technique \cite{Haldane81}, is available
\cite{CazalillaHo03}. However, the approach of Ref.\ \cite{CazalillaHo03} is constrained by the requirement
that the inter-species coupling is small compared to the characteristic
bandwidth, which becomes too restrictive for dilute systems and
leads to a breakdown of bosonization at very low densities.

The goal of the present work is to provide a stability analysis
for dilute multicomponent Bose gases in low dimensions, valid for any coupling
strength. It is shown that  strong renormalization of coupling constants,
typical for low-dimensional systems, can lead to a substantial change in the
stability conditions compared to mean-field results, particularly in the case of  inequivalent
species (heteronuclear mixtures); the mean-field answers restore their validity
with the increase of density. The outline of the paper is as follows: in Section
\ref{sec:RG} I present the generalization of the renormalization group (RG)
approach to the multicomponent case, in Section \ref{sec:2comp} I apply this
general formalism to the simplest case of two bosonic species with
density-density interaction, and Section \ref{sec:F1} illustrates the case of
mutually convertible species on the example of spin-one bosons. Finally, Section
\ref{sec:summary} contains the discussion of the possible numerical and physical
tests of the theory predictions.

\section{Renormalization group approach for a multicomponent dilute Bose gas}
\label{sec:RG}

As a starting point for our analysis, we will choose the theory of
multi-species bosonic field with pointlike two-body interactions, that
effectively describes a system of atoms with typical momenta much
smaller than the inverse characteristic potential range (atom size).
Effective field theory in combination with the renormalization group
analysis is a convenient tool
\cite{Andersen04,BraatenHammer06,Braaten+08} that is widely used
\cite{BijlsmaStoof96,BraatenNieto97,NikolicSachdev07} for the
study of the cold atom systems.

Consider the following continuum action describing $N$ species of bosons 
at zero temperature in $d$ spatial dimensions, coupled by a general quartic contact interaction:
\begin{eqnarray} 
\label{N-action} 
\mathcal{A}_{N} &=& \int d\tau \int d^{d}x \Big\{  
\psi_{\alpha}^{*}(\partial_{\tau}-\mu_{\alpha})\psi_{\alpha}+\frac{|\nabla
  \psi_{\alpha}|^{2}}{2m_{\alpha}} +U\Big\}, \nonumber\\
U&=&\frac{1}{2}\sum_{\alpha\beta\alpha'\beta'}
g_{\alpha\beta,\alpha'\beta'}
\psi^{*}_{\alpha}\psi^{*}_{\beta}\psi_{\alpha'}^{\vphantom{*}}\psi_{\beta'}^{\vphantom{*}}.
\end{eqnarray}
Here $\psi_{\alpha}$ are the fields describing bosonic particles with masses
$m_{\alpha}$, $\alpha=1,\ldots N$, and we have set $\hbar=1$. The interaction
matrix for Bose fields satisfies the obvious symmetry  conditions
$g_{\alpha\beta,\alpha'\beta'}=g_{\beta\alpha,\alpha'\beta'}
=g_{\alpha\beta,\beta'\alpha'}=g^{*}_{\alpha'\beta',\alpha\beta}$.  
The action (\ref{N-action}) can  describe a mixture in a continuum as well as in
an optical lattice \cite{CazalillaHo03}.

\begin{figure}[tb]
\begin{center}
\includegraphics[width=0.25\textwidth]{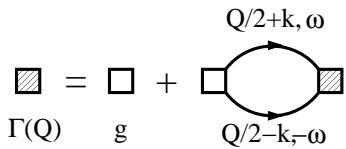}
\end{center}
\caption{\label{fig:vertex} The renormalized interaction vertex for the multicomponent Bose
  gas model
  (\ref{N-action}) in the zero density limit, see Eq.\ (\ref{vertex}).
}
\end{figure}

When all
chemical potentials $\mu_{\alpha}$ vanish, the system is at a critical point. At
this special point, the physical picture is considerably simplified \cite{Uzunov81,Fisher+89}:
there is no self-energy correction so that the full
propagator just coincides with the free one $G_{\alpha\beta}(k,\omega)
=\delta_{\alpha\beta}/(i\omega-\varepsilon^{(\alpha)}_{k})$, where
$\varepsilon^{(\alpha)}_{k}=k^{2}/2m_{\alpha}$. This is easy to understand
physically, since when all
$\mu_{\alpha}=0$, the particle density is just zero. Interaction between the
particles is, however,
 renormalized due to multiple scatterings. 
The renormalized
vertex $\Gamma$, defined at the fixed sum of external momenta $Q$, satisfies the following Bethe-Salpeter equation
(illustrated in Fig.\ \ref{fig:vertex})
\begin{eqnarray} 
\label{vertex} 
\Gamma_{\alpha\beta,\gamma\delta}(Q)&=&g_{\alpha\beta,\gamma\delta}
-g_{\alpha\beta,\alpha'\beta'}f_{\alpha'\beta'}(Q)\Gamma_{\alpha'\beta',\gamma\delta}(Q),\nonumber\\
f_{\alpha\beta}(Q)
&=&
\int
\frac{d^{d}k}{(2\pi)^{d}}
\big[\varepsilon^{(\alpha)}_{Q/2+k} + \varepsilon^{(\beta)}_{Q/2-k}\big]^{-1}.
\end{eqnarray}
The simple ladder form of the above equation is due to the fact that
contributions from any diagrams containing closed loops vanish \cite{Uzunov81}.

 Above the upper critical dimension $d=2$, $f_{\alpha\beta}(Q)$ converges at
 $Q\to 0$ and (\ref{vertex}) simply yields the dressed (observable) interaction
 matrix.  For $d\leq 2$ there is a singularity at $Q\to 0$:
 $f_{\alpha\beta}(Q)\propto 1/Q$ for $d=1$ and $f_{\alpha\beta}(Q)\propto
 \ln(\Lambda_{0}/Q)$ for $d=2$, where $\Lambda_{0}$ is the ultraviolet cutoff
 (for a system in an optical lattice, the parameter $\Lambda_{0}$ has the
 physical meaning of a lattice cutoff, and in a continuum it has the sense of an
 inverse characteristic potential range).
Thus for $d\leq 2$ one should look at the
RG flow of the
running coupling matrix $\Gamma(l)=\Gamma(Q) e^{(2-d)l}$ with the change of the scale $Q\mapsto
\Lambda_{0}e^{-l}$.  Defining the matrices 
\begin{eqnarray} 
\label{hilfematrizen} 
&&\overline{\Gamma}_{\alpha\beta,\alpha'\beta'}=2\Gamma_{\alpha\beta,\alpha'\beta'}
(m_{\alpha\beta}m_{\alpha'\beta'})^{1/2},\nonumber\\
&&F_{\alpha\beta,\alpha'\beta'}(Q)=\delta_{\alpha\beta,\alpha'\beta'}R_{d}(Q_{\alpha\beta}),
\end{eqnarray} 
where $m_{\alpha\beta}\equiv m_{\alpha}m_{\beta}/(m_{\alpha}+m_{\beta})$ are
the reduced masses,
$Q_{\alpha\beta}=2Q(m_{\alpha}m_{\beta})^{1/2}/(m_{\alpha}+m_{\beta})$, and
\[
R_{d}(q)=\int \frac{d^{d}k}{(2\pi)^{d}}\frac{1}{k^{2}+q^{2}}\to 
\begin{cases}
\frac{1}{2q}, & d=1\\
\frac{1}{2\pi}\ln\frac{\Lambda_{0}}{q}, & d=2
\end{cases}
,
\]
one can write down the RG equation as
\begin{equation} 
\label{RG-gen-a} 
\frac{d}{dl}\Big\{ \overline{\Gamma}(1-F\overline{\Gamma})^{-1}
\Big\}=0.
\end{equation}
Further, using the identity
$AX = XA = X-1$ with  $X\equiv (1-A)^{-1}$ and its derivative $dX/dl=X(dA/dl)X$,
those equations can be rewritten in the following form:
\begin{equation} 
\label{RG-gen-b} 
d\overline{\Gamma}/dl+ \overline{\Gamma}(dF/dl)\overline{\Gamma} =0.
\end{equation}
For further simplification,
it is convenient to redefine the
interaction matrix once again.
In one dimension ($d=1$), one can introduce
$\widetilde{\Gamma}=F^{1/2}\overline{\Gamma}F^{1/2}$; then,  since $F$ is a
diagonal matrix that for $d=1$
is proportional to $1/Q$, one obtains $dF/dl=F$ and  Eq.\ (\ref{RG-gen-b}) can
be rewritten as
\begin{equation}
\label{RG-1d-b} 
d\widetilde{\Gamma}/dl=\widetilde{\Gamma} -\widetilde{\Gamma}^{2},
\end{equation}
which is familiar from the one-component case \cite{Sachdev-book}; the only
difference is that the interaction is now a matrix. 

In two dimensions ($d=2$), it is convenient to define
$\widetilde{\Gamma}=\overline{\Gamma}/2\pi$; then, using the fact that $dF/dl=1/(2\pi)$ for $d=2$,
one can reduce (\ref{RG-gen-b}) to 
\begin{equation} 
\label{Rg-2d} 
d\widetilde{\Gamma}/dl= -\widetilde{\Gamma}^{2}.
\end{equation}

Summarizing the above derivation, we see that the RG equations for the
interaction matrix of the low-dimensional multicomponent Bose gas can be cast into the common form
\begin{equation} 
\label{RG} 
d\widetilde{\Gamma}/dl=(2-d)\widetilde{\Gamma} -\widetilde{\Gamma}^{2},
\end{equation}
where 
\begin{eqnarray} 
\label{matrix} 
\widetilde{\Gamma}_{\alpha\beta,\gamma\delta}(l)&=&
\Gamma_{\alpha\beta,\gamma\delta}(l)
(m_{\alpha}m_{\beta}m_{\gamma}m_{\delta})^{1/4} /\Lambda_{0}
,\quad
d=1, \nonumber\\
\widetilde{\Gamma}_{\alpha\beta,\gamma\delta}(l)&=&
\Gamma_{\alpha\beta,\gamma\delta}(l)
\sqrt{m_{\alpha\beta}m_{\gamma\delta}}/\pi
,\quad d=2.
\end{eqnarray}
Since the
$\widetilde{\Gamma}$   matrices are symmetric, they can always be diagonalized by an
appropriate orthogonal transformation.  
It is worth noting that \emph{at the zero-density critical point} the equations (\ref{RG}) are exact to all
orders in the interaction $\widetilde{\Gamma}$, similar to the one-component
case \cite{Sachdev-book}. 

The above derivation is nothing but a direct generalization of the well-known RG approach to
the \emph{one-component} dilute Bose gas
\cite{FisherHohenberg88,NelsonSeung89,Fisher+89,KolomeiskyStraley92}.
 Similar to the one-component case, we will use the RG equations
 (\ref{RG}), derived at the critical point (zero particle density), to describe
 the gas of a \emph{finite} but small density, and to stop the RG flow at some
 scale $l=l^{*}$ where the  system is effectively no more dilute, i.e., where the ``running'' total density
$\rho(l)=\rho_{\rm tot}e^{dl}$ becomes comparable with $\Lambda_{0}^{d}$. This
 condition determines the stopping scale $l^{*}$ as
\begin{equation} 
\label{l*} 
e^{dl^{*}} =C_{d}\Lambda_{0}^{d}/\rho_{\rm tot},
\end{equation}
where $C_{d=1}=\frac{2}{\pi^{2}}$ and $C_{d=2}=\frac{1}{2\pi}$ can be identified
by comparing to the known results for the one-component case \cite{KolomeiskyStraley92}.

\section{Two-component Bose mixture}
\label{sec:2comp}

Let us apply the RG approach outlined in the previous section to the simplest
case of two non-convertible species (i.e., the particle numbers
are separtely conserved for each species). 
The quartic interaction in (\ref{N-action})
then takes the form
\begin{equation} 
\label{U-2comp} 
U(\psi_{1},\psi_{2})=(u_{11}|\psi_{1}|^{4} 
+u_{22}|\psi_{2}|^{4})/2 +u_{12}|\psi_{1}|^{2}|\psi_{2}|^{2},
\end{equation}
so the bare couplings are $g_{11,11}=u_{11}$, $g_{22,22}=u_{22}$, and $g_{12,12}=u_{12}/2$. If we
denote the  entries of the $\widetilde{\Gamma}$ matrix by
$\widetilde{u}_{11}$, etc., the RG equations (\ref{RG}) will simplify to
\begin{equation} 
\label{RG-2comp} 
d\widetilde{x}/dl=(2-d)\widetilde{x}- \widetilde{x}^{2},
\end{equation}
with $\widetilde{x}\in \{\widetilde{u}_{\alpha\beta} \}$.
We focus on the stability of the mixture. The potential 
(\ref{U-2comp}) is a quadratic form of the densities, so the \emph{necessary} condition
of the local stability is positiveness of its discriminant:
\begin{equation} 
\label{discrim} 
u_{11}(l)u_{22}(l)-u_{12}^{2}(l)>0,
\end{equation}
which must be satisfied by the renormalized interactions $u_{\alpha\beta}(l)$ at
all scales up to the RG stopping scale $l=l^{*}$; the intra-species couplings
$u_{\alpha\alpha}$ must also be assumed positive to ensure stability. Violation
of the condition (\ref{discrim}) leads to phase separation (demixing transition)
for repulsive interactions between species ($u_{12}>0$) and to a collapse in
case of attraction ($u_{12}<0$). At the level of bare coupling 
($l=0$), Eq.\ (\ref{discrim}) corresponds to the stability condition obtained in
the mean-field analysis \cite{HoShenoy96,Timmermans98}.

\subsection{$d=1$}

In the case of one spatial dimension the solution of RG equations
(\ref{RG-2comp}) reads
\begin{eqnarray} 
\label{d1sol} 
&& \widetilde{u}_{\alpha\beta}(l)=\Big\{1+\big(1/\widetilde{u}^{(0)}_{\alpha\beta}-1\big)
e^{-l}\Big\}^{-1}, \nonumber\\
&&\widetilde{u}_{\alpha\beta}(l)= u_{\alpha\beta}(l) \sqrt{m_{\alpha}m_{\beta}}/\Lambda_{0} ,
\end{eqnarray}
where $\widetilde{u}^{(0)}_{\alpha\beta}$ denotes the corresponding bare ($l=0$)
value. A peculiarity of the 1D case is the tendency of the RG flow to establish an
enhanced symmetry if the interaction between the species is repulsive
($u_{12}>0$), \emph{even for inequivalent species}. Indeed, one can see that at the fixed point 
$\widetilde{u}_{\alpha\beta}=1$ the potential
(\ref{U-2comp}) tends to
\[
U(\psi_{1},\psi_{2})\mapsto 
\frac{\Lambda_{0}}{2}\Big(\frac{|\psi_{1}|^{2}}{\sqrt{m_{1}}}+\frac{|\psi_{2}|^{2}}{\sqrt{m_{2}}} \Big)^{2},
\]
so the original $U(1)\times U(1)$ symmetry is enhanced to $U(1) \times SU(2)$ at
the fixed point.

\begin{figure}[tb]
\begin{center}
\includegraphics[width=0.3\textwidth]{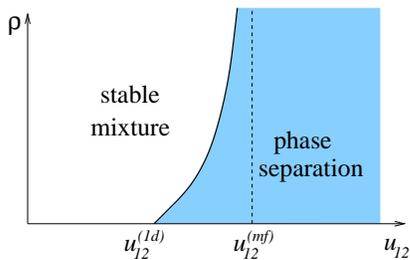}
\end{center}
\caption{\label{fig:phd-1d} (Color online). The phase diagram of a
  two-component one-dimensional dilute Bose mixture. The behavior of
  the critical density $\rho$ in the vicinity of $u_{12}=u_{12}^{(1d)}$
  is linear, cf. Eq.\ (\ref{d1-rhoc}). 
}
\end{figure}

Let us first consider the repulsive case.
Assume that at the microscopic scale ($l=0$) the mixture is stable, i.e.,
$u_{12}^{2}<u_{11}u_{22}$. Then the  stability condition (\ref{discrim}) breaks down
starting from a certain scale $l=l_{c}$, determined by
\begin{eqnarray} 
\label{d1-lc} 
&& e^{l_{c}}=1+
\frac{\Lambda_{0}\big\{[u_{12}^{(mf)}]^{2} -u_{12}^{2}\big\} }
{u_{12}(u_{11}m_{1}+u_{22}m_{2})\big\{u_{12}-u_{12}^{(1d)} \big\} },\\
&& u_{12}^{(1d)}=\frac{2u_{11}u_{22}\sqrt{m_{1}m_{2}}}{u_{11}m_{1}+u_{22}m_{2}},
\quad u_{12}^{(mf)}=\sqrt{u_{11}u_{22}}.\nonumber
\end{eqnarray}
The instability develops if $l^{*}>l_{c}$, which 
translates into the requirement for the density to be low enough,
\begin{equation} 
\label{dilute} 
\rho_{\rm tot}<C_{1}\Lambda_{0}e^{-l_{c}}.
\end{equation}
Note that $l_{c}$ is \emph{not} replacing the RG stopping scale $l^{*}$, but is 
a new scale that naturally arises in the problem. Here $u_{12}^{(mf)}$ denotes
the mean-field result for the instability point \cite{HoShenoy96,Timmermans98}, 
but the actual
instability, as it is easy to see, generally occurs already at smaller values of $u_{12}$.
Indeed, 
Eq.\ (\ref{d1-lc}) has a real solution $l_{c}$  if $u_{12}^{(1d)}< u_{12}<
u_{12}^{(mf)}$, and one has  $l_{c}\to\infty (0)$ as $u_{12}\to u_{12}^{(1d)}
(u_{12}^{(mf)})$. 
 If the ultraviolet cutoff
$\Lambda_{0}$ is much larger than all the other energy scales, the 
critical density $\rho_{c}$, below which the instability develops
for a given $u_{12}$,
becomes cutoff-independent and is given by
\begin{equation}
\label{d1-rhoc}
\rho_{c} =
 \frac{C_{1} u_{12}(u_{11}m_{1}+u_{22}m_{2})\big\{ u_{12}-u_{12}^{(1d)}\big\}}{[u_{12}^{(mf)}]^{2} -u_{12}^{2} }.
\end{equation}
The above formula can be alternatively viewed as determining 
the critical value of the interaction $u_{12}^{(c)}$ for a  given
density $\rho_{\rm tot}=\rho_{c}$. One can see that
the actual instability point $u_{12}^{(c)} \to u_{12}^{(1d)}$ if $\rho_{\rm tot}\to
0$, and tends to the mean-field answer $u_{12}^{(c)}\mapsto u_{12}^{(mf)}$ when
the density becomes large. 
The resulting phase diagram for the one-dimensional case is schematically shown
in Fig.\ \ref{fig:phd-1d}.
For equivalent species ($u_{11}=u_{22}$, $m_{1}=m_{2}$), the points
$u_{12}^{(1d)}$ and $u_{12}^{(mf)}$ coincide, and there is no change in the
stability condition compared to the mean-field case.

The case of attractive interspecies interactions ($u_{12}<0$) is much more
complicated. The RG flow for $u_{12}(l)$ exhibits a runaway singularity at the characteristic scale
$l_{b}\simeq\ln\frac{\Lambda_{0}}{|u_{12}|\sqrt{m_{1}m_{2}}}$,
namely, $u_{12}(l)\to -\infty$ as $l\to l_{b}$. It is easy to see that this
scale is connected to the presence of molecules (bound states), which exist in
one dimension at
any strength of the attraction $u_{12}$. Indeed, the energy of the bound
state for the contact interaction $U(x)=-|u_{12}|\delta(x)$ is $E_{b}=-\frac{1}{2} m_{12}u_{12}^{2}$, and
the total energy of a  molecule with the total momentum $Q$ is $E_{\rm
  mol}(Q)=E_{b}+Q^{2}/2(m_{1}+m_{2})$; molecules with momenta
$|Q|>Q_{b}=\Lambda_{0}e^{-l_{b}}$ can be broken up in collisions, while those
with $|Q|<Q_{b}$ are stable. 

Molecules represent a new ``species'', so it is necessary to introduce a new molecular bosonic field $\Psi_{m}$, and in addition
to atom-molecule and molecule-molecule interaction one has to include three-particle amplitudes of the type
$\Psi_{m}\psi_{1}^{\dag}\psi_{2}^{\dag}$. Presence of those amplitudes 
invalidates the initial model (\ref{N-action}), so the
attractive case requires a separate analysis which is beyond the scope of the
present paper.

I would like to conclude this subsection with a remark concerning the relation
between the present study and that by
Cazalilla and Ho \cite{CazalillaHo03}. On the basis of the bosonization
analysis, they
have  pointed out  that the stability condition is
of the mean-field type for large densities (quasicondensate regime) but
changes at low densities. Their instability criterion for the low
density case (which corresponds to the Tonks limit
$m_{\alpha}u_{\alpha\alpha}/\rho_{\alpha}\gg 1$, here $\rho_{\alpha}$ are
the densities of individual species) reads 
\begin{equation} 
\label{cazalilla} 
|u_{12}|>\pi^{2}(\rho_{1}\rho_{2}/m_{1}m_{2})^{1/2},
\end{equation}
and thus is different from that
derived in the present work. However, one can argue that the condition (\ref{cazalilla})
should be rather taken as
the  applicability limit of the bosonization approach itself. Indeed, the
perturbative approach of  \cite{CazalillaHo03} can be expected to break down
when the
inter-species interaction $u_{12}$ becomes comparable with the characteristic
bandwidths $v_{\alpha}=\pi\rho_{\alpha}/m_{\alpha}$  of the individual species. 
One can also disprove the criterion (\ref{cazalilla}) by 
considering the limit of hardcore bosons $u_{11}, u_{22}\to\infty$ on
a lattice: in this limit our model is formally equivalent to that of two
fermionic species with an on-site interaction, i.e., to the one-dimensional
Hubbard model. This model is exactly solvable and from its phase diagram it is
known \cite{1DHubbard-book} that there is no instability for arbitrary small
densities. At the same time,
 the lower
 boundary for the instability point $u_{12}^{(1d)}$, given by (\ref{d1-lc}),
 diverges as $u_{\alpha\alpha}\to \infty$, so  the absence of instability in the
 hardcore limit is correctly reproduced in the present approach.

\subsection{$d=2$}

In the two-dimensional case  the solutions of RG equations are given by
\begin{equation} 
\label{d2sol} 
 \widetilde{u}_{\alpha\beta}(l)=\big\{ l+1/\widetilde{u}_{\alpha\beta}^{(0)}
\big\}^{-1}, 
\quad
\widetilde{u}_{\alpha\beta}(l)=\frac{m_{\alpha\beta}}{\pi} u_{\alpha\beta}(l).
\end{equation}
In contrast to the one-dimensional case, a tendency to the
enhanced $SU(2)$ symmetry is only present for species of equal masses.

\begin{figure}[tb]
\begin{center}
\includegraphics[height=0.2\textwidth]{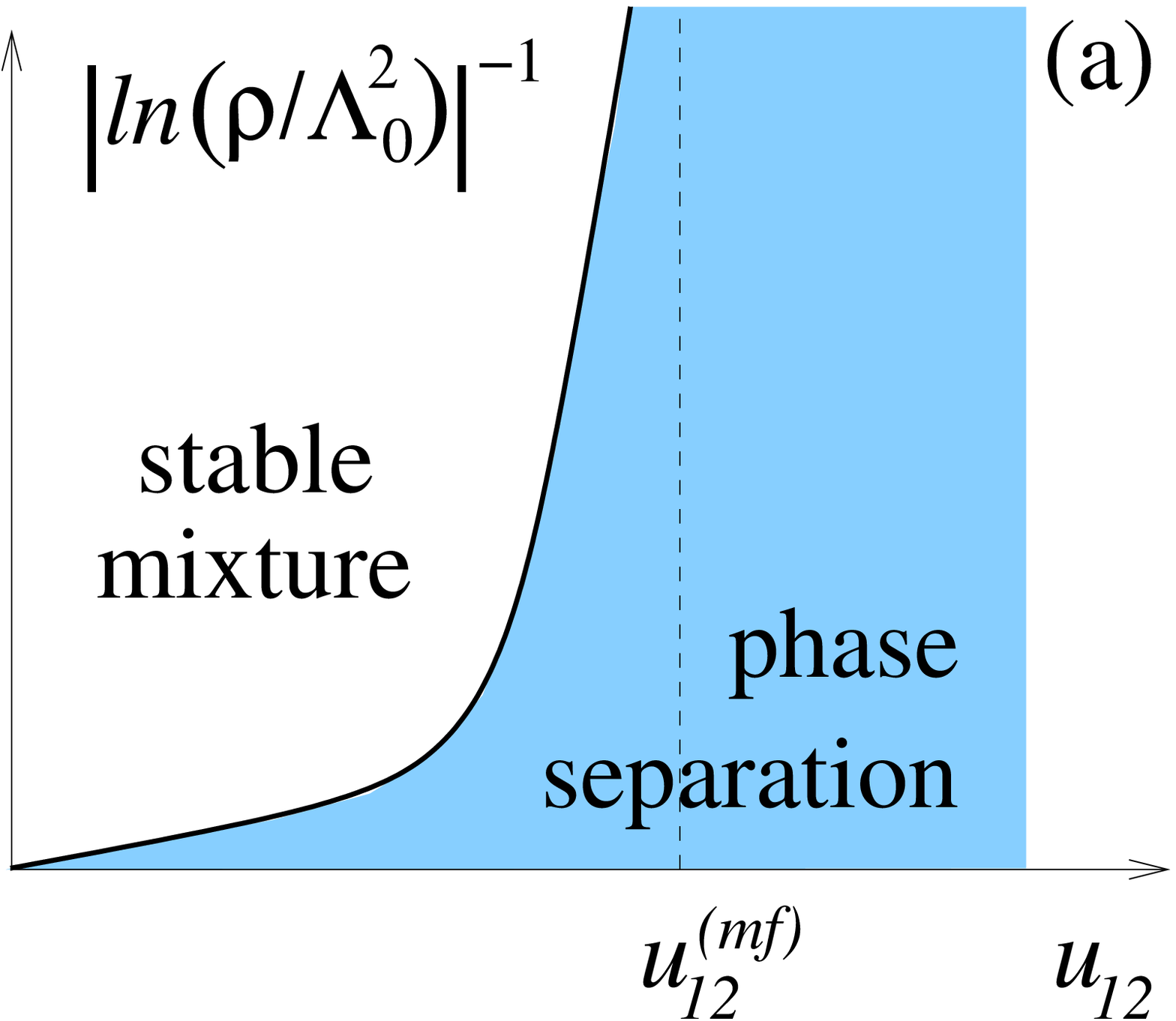}\hspace*{2mm}
\includegraphics[height=0.2\textwidth]{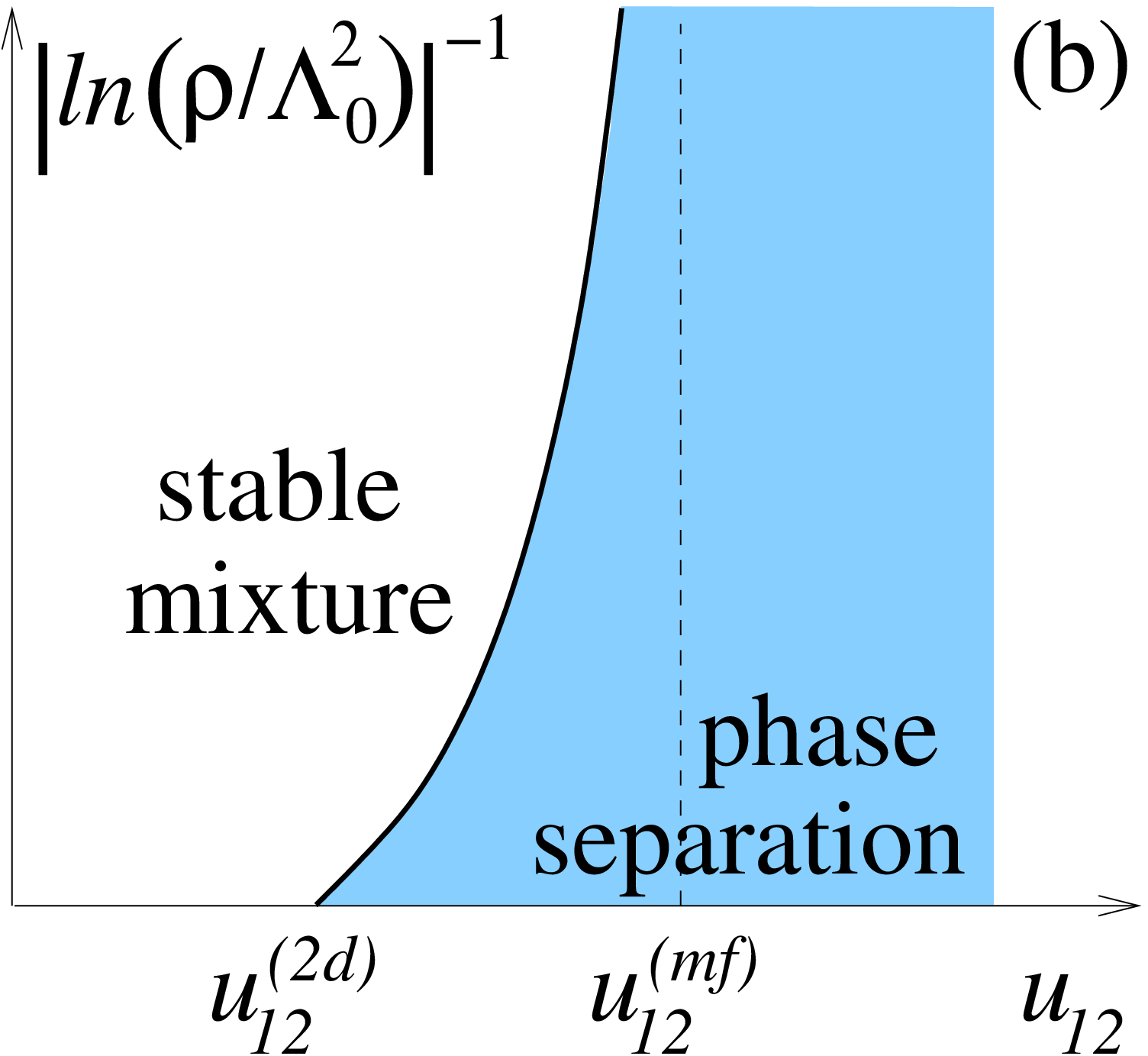}
\end{center}
\caption{\label{fig:phd-2d} (Color online) The phase diagram of a
  two-component dilute Bose mixture in two dimensions: (a) the general case; (b)
  the special case of equal species masses $m_{1}=m_{2}$. The behavior of
  the critical values of 
  $1/|\ln(\rho/\Lambda_{0}^{2})|$ in the vicinity of $u_{12}=0$ and
  $u_{12}=u_{12}^{(2d)}$
  is linear, cf. Eqs.\ (\ref{d2-lc}), (\ref{d2-rhoc}).
}
\end{figure}

Following a similar route as for $d=1$, one obtains the scale $l_{c}$ at which
the  stability condition (\ref{discrim}) breaks down: it is given by the positive root of the equation
\begin{eqnarray} 
\label{d2-lc} 
&&\Big(\frac{m_{1}-m_{2}}{m_{1}+m_{2}}\Big)^{2} l^{2} 
-\frac{4\pi^{2}}{m_{1}m_{2}}\Big\{ \frac{1}{u_{12}^{2}}-\frac{1}{u_{11}u_{22}} \Big\}\\
&&\qquad +2\pi l\Big\{\frac{1}{u_{11}m_{1}} +\frac{1}{u_{22}m_{2}}
-\frac{4}{u_{12}(m_{1}+m_{2})} \Big\}  =0.\nonumber
\end{eqnarray}
In the general case of unequal masses ($m_{1}\not=m_{2}$), the solution $l_{c}$ exists for
any $u_{12}$ within the interval $-u_{12}^{(mf)} < u_{12} < u_{12}^{(mf)}$; the
scale $l_{c}$ vanishes for
$u_{12}\to\pm u_{12}^{(mf)}$ and diverges at $u_{12}\to 0$. 
An instability occurs if $l_{c}<l^{*}$, i.e., if the total density is low enough to
satisfy the inequality 
\begin{equation} 
\label{d2-rhoc} 
\rho_{\rm tot}<C_{2}\Lambda_{0}^{2}e^{-2l_{c}}.
\end{equation} 
This means that, in
contrast to $d=1$, in two dimensions a dilute Bose mixture of two \emph{unequal} species is
always unstable below a certain density, for \emph{any} value of the
interspecies interaction $u_{12}$, see Fig.\ \ref{fig:phd-2d}(a). 
However, for small $u_{12}$ the critical density is exponentially
small: according to Eqs.\ (\ref{d1-lc}), (\ref{d1-rhoc}), at
$u_{12}\to +0$ the critical value of the density behaves as 
$\rho_{c}\simeq C_{2}\Lambda_{0}^{2} \exp(-A/u_{12})$, where
\begin{equation} 
\label{d2-u0} 
A=\frac{4\pi (m_{1}+m_{2})}{\sqrt{m_{1}m_{2}}(\sqrt{m_{1}}- \sqrt{m_{2}})^{2}}.
\end{equation}
At $m_{1}\to m_{2}$ the coefficient $A$ diverges;  actually, in that limit
the critical density rapidly goes to zero in the whole interval $0<u_{12}<u_{12}^{(2d)}$,
where
\begin{equation} 
\label{u2d} 
u_{12}^{(2d)}=\frac{2u_{11}u_{22}}{u_{11}+u_{22}}.
\end{equation}
Thus, the case of equal masses is special: $l_{c}$ exists only for
$u_{12}^{(2d)}<u_{12}^{(mf)}$, resembling the situation in the one-dimensional
case, see Fig.\ \ref{fig:phd-2d}(b). If not only the masses, but also the
intraspecies couplings of the two species are equal ($u_{11}=u_{22}$), the
mean-field answer for the instability threshold is restored at all densities.

The case of attractive interspecies interaction  suffers from a similar runaway
singularity as in the 1D case. The scale $l_{b}=\pi(m_{1}+m_{2})/m_{1}m_{2}|u_{12}|$,
at which the running coupling $u_{12}(l)$ diverges, is connected to the
existence of a bound state (molecule) with the binding energy $E_{b}\propto
\exp(-2l_{b})$. Presence of molecules invalidates our original framework
(\ref{N-action}), so it cannot be directly applied to the attractive case.

\section{$F=1$ spinor bosons} 
\label{sec:F1}

In this section I would like to illustrate the application of the presented
method to a system with convertible species, where the total particle number of
individual species is not conserved.
Consider a gas of bosons with the hyperfine spin
$F=1$. The contact interaction between two paticles depends on their total spin $S$, so
there are two characteristic scattering lengths
$a_{S}$ with $S=0$, $2$, and the interaction can be written as \cite{Ho98,OhmiMachida98}
\begin{equation} 
\label{U-f1} 
U=\frac{1}{2}\big\{ c_{0} 
\psi_{a}^{\dag}\psi_{a'}^{\dag}
\psi_{a'}^{\vphantom{\dag}}\psi_{a}^{\vphantom{\dag}} 
+ c_{2}\psi_{a}^{\dag}\psi_{a'}^{\dag} F^{\mu}_{ab} F^{\mu}_{a'b'}\psi_{b} \psi_{b'}
 - h(S^{z})^{2}\big\},
\end{equation}
where $a,b\in\{0,\pm1\}$ denote the three components of the bosonic field,
$F^{\mu}$ are the spin-$1$ matrices, and
$S^{\mu}=\psi_{a}^{\dag}F^{\mu}_{ab}\psi_{b}$ are the spin operators.  The
couplings are given by $c_{0}= (g_{0}+2g_{2})/3$, $c_{2}=(g_{2}-g_{0})/3$, with
$g_{S} \propto a_{S}$, and the mass $m$ is the same for all three components.
The last term in (\ref{U-f1}) describes the quadratic Zeeman effect caused by an
external magnetic field.  This type of interaction favors ferromagnetic spin
correlations for $c_{2}<0$, and polar (nematic) ones for $c_{2}>0$
\cite{Ho98,OhmiMachida98}.

\begin{figure}[tb]
\begin{center}
\includegraphics[width=0.34\textwidth]{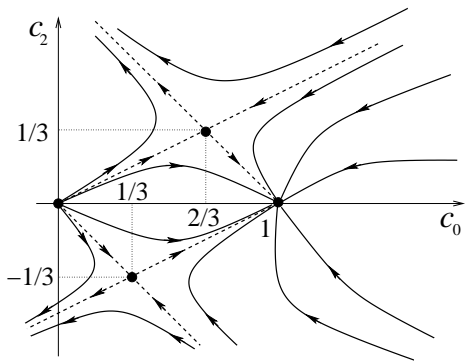}
\end{center}
\caption{\label{fig:rg-f1h0} RG flow diagram for 
the spinor boson model (\ref{U-f1}) at zero magnetic field $h=0$, for the
one-dimensional case.
}
\end{figure}

Although
(\ref{U-f1}) includes just three parameters, the $9\times9$ interaction matrix
$g_{ab,a'b'}$ contains generally five different couplings:
\begin{eqnarray}
\label{f1-couplings}
g_{++,++} = g_{--,--} &=& r \mapsto c_{0}+c_{2}-h \nonumber\\
2g_{+0,+0} = 2g_{-0,-0} &=& f \mapsto c_{0}+c_{2} \nonumber\\
 2g_{+-,+-} &=& v \mapsto c_{0}-c_{2}+h,\\
g_{00,00} &=& u \mapsto c_{0},\quad
g_{+-,00} =w \mapsto 2c_{2}, \nonumber
\end{eqnarray}
where arrows indicate the corresponding bare values.

At zero external field ($h=0$) the
$SU(2)$ symmetry dictates that the effective potential can contain only two
constants $c_{0}$ and $c_{2}$, and the RG equations can be cast into the form (\ref{RG-2comp}) with $ \widetilde{x}\in \{
\widetilde{c}_{0}+\widetilde{c}_{2},
\widetilde{c}_{0}-2\widetilde{c}_{2}\}$. 
The ``tilded'' variables are here defined as
$\widetilde{x}=xm/\Lambda_{0}$ for $d=1$ and $\widetilde{x}=xm/2\pi$ for $d=2$.
In one dimension ($d=1$), there is a
nontrivial stable fixed point $(\widetilde{c}_{0}=1,\widetilde{c}_{2}=0)$, which
exhibits the enhanced $SU(3)$ symmetry, 
similar to the double-species case where the fixed point was
$SU(2)$-symmetric; the other two fixed points,
$(\widetilde{c}_{0}=2/3,\widetilde{c}_{2}=1/3)$ and
$(\widetilde{c}_{0}=1/3,\widetilde{c}_{2}=-1/3)$ are unstable.
The corresponding RG flow for $d=1$ is shown in Fig.\ \ref{fig:rg-f1h0}.
Both for  $d=1$ and $d=2$, there are two runaway flows: the one 
at $c_{2}>c_{0}/2$ marks the onset of the pairing state
 characterized by the
formation of bound singlet pairs \cite{CaoJiangWang07,Essler+09,KunYang09}, while the other runaway flow at
$c_{2}<-c_{0}$ ($c_{0}>0$) corresponds to the collapse instability when all atoms tend to
bind into one giant ferromagnetic ``drop''. Except of those two runaways, there
are no other instabilities.

In presence of the quadratic Zeeman term ($h\not=0$) one cannot simply use the
full matrix (\ref{f1-couplings}) in the RG equations, because a nonzero $h$
leads to a relative shift in
chemical potentials of different species, suppressing the  $|0\rangle$ states for $h>0$ and
$|\pm1\rangle$ states for $h<0$, respectively. Therefore, if $|h|$ is large
enough, one can neglect the suppressed species. For instance, for $h>0$  one comes back
to a double-species problem for $|+\rangle$ and $|-\rangle$ states, with a phase separation occuring at $v>r$, $r>0$,
i.e., at $c_{0}+c_{2}>h>c_{2}$. At $c_{2}>0$ this phase separation corresponds
to a transition between the nematic and ferromagnetic states. 
For $h<0$ one effectively obtains a one-component gas, so nothing interesting happens. 

\section{Summary and discussion} 
\label{sec:summary}

The examples considered above show that a straightforward generalization of the
renormalization group approach for dilute low-dimensional Bose gas
\cite{FisherHohenberg88,Fisher+89,KolomeiskyStraley92} to the case of
multicomponent systems allows one to obtain nontrivial results concerning the
stability of Bose mixtures. For inequivalent species, the stability conditions
deviate strongly from the mean-field results; the latter are recovered at higher
density.

It is worth
noting that the  action (\ref{N-action}) is quite general: 
except being applicable to a multicomponent gas in a continuum as well
as in an optical lattice, it also
arises in  quantum magnetism problems
dealing with the so-called Bose-Einstein condensation of magnons, induced by a
strong external magnetic field $H$ in the vicinity of the saturation field
$H_{s}$ 
\cite{BatyevBraginskii84}. In a 
frustrated magnet the magnon dispersion may have degenerate minima
at inequivalent wave vectors, which gives rise to multiple ``species'' in an
effective model \cite{NikuniShiba95,JackeliZhitomirsky04,KolezhukVekua05}, 
while  $H_{s}-H$ plays the role of the chemical potential.

It would be interesting to test the predictions of the present paper numerically
for a one-dimensional two-component Bose-Hubbard model with on-site
interaction, described by the Hamiltonian
\begin{eqnarray} 
\label{bh-ham} 
\mathcal{H}&=&-\sum_{j,\alpha}t_{\alpha}(b^{\dag}_{j,\alpha}b^{\vphantom{\dag}}_{j+1,\alpha}
+\mbox{h.c.})\\
&+&\frac{1}{2}\sum_{j,\alpha}U_{\alpha\alpha}\widehat{n}_{j,\alpha}(\widehat{n}_{j,\alpha}-1)
+U_{12}\sum_{j}\widehat{n}_{j,1}\widehat{n}_{j,2}, \nonumber
\end{eqnarray}
where $\alpha=1,2$ labels the two species, $b_{j,\alpha}$ is a bosonic operator
at the lattice site $j$, and
$\widehat{n}_{j,\alpha}=b^{\dag}_{j,\alpha}b^{\vphantom{\dag}}_{j,\alpha}$ are
the corresponding number operators.
A simple estimate shows that the necessary requirements can be
satisfied at realistic values of the model parameters. Taking, for example, two
species with different hopping amplitudes $t_{1}=t$, $t_{2}=t/5$ and the same
intra-species repulsion $U_{11}=U_{22}=U=3t$, from (\ref{d1-lc}) one obtains the
following window in the inter-species coupling where the phase separation can
take place: $0.745 < U_{12}/U < 1$. Further, fixing, for instance, $U_{12}=0.8U$, we find from
(\ref{dilute}) that, in order for the instability to develop, the total density
per site $n=\rho_{\rm tot}/\Lambda_{0}$ must satisfy $n<0.16$, which can be
reasonably reached in a density matrix renormalization group calculation
\cite{Kleine+08}.

Experimentally, the most promising candidate is the
$\rm{}^{87}Rb$-$\rm{}^{41}K$ mixture, for which a very fine control of
interactions has been achieved \cite{Thalhammer+08}. Assuming that all atoms are
in their $|1,1\rangle$ state (as in the setup of Ref.\ \cite{Thalhammer+08}), at low magnetic field $B<100$~G the
intra-species scattering lengths are approximately given by their $B=0$ values 
$a_{11}\simeq 97.4$  for $\rm{}^{87}Rb$ \cite{vanKempen+02} and
$a_{22}\simeq 65.2$ for $\rm{}^{41}K$ \cite{Falke+08}, in units of 
the Bohr radius. The effective couplings in a cigar-shaped (quasi-1D) trap with  the
transverse confining frequency $\omega_{\perp}$ are given by
$u_{\alpha\beta}=2\hbar \omega_{\perp}a_{\alpha\beta}$
 \cite{Olshanii98} (where it is assumed that the characteristic confinement
 radius $a_{\perp}\gg a_{\alpha\beta}$). 
Thus,
from Eq.\ (\ref{d1-lc}) the critical values for the inter-species
scattering length are $a_{12}^{(1d)}\simeq 68.05$ and $a_{12}^{(mf)}\simeq
79.7$ (independent of $\omega_{\perp}$). According to Ref.\ \cite{Thalhammer+08},
  the dependence of the interspecies
scattering length $a_{12}$ on the magnetic field $B$ is very well  described by the
following expression obtained in Ref.\ \cite{Simoni+08}:
\begin{equation} 
\label{a12}
a_{12}(B)=a_{bg}\sum_{n=1,2}\Delta_{n}/(B-B_{n}),  
\end{equation}
with $a_{bg}=284$~a.u., $\Delta_{1}=37$~G, $B_{1}=39.4$~G, $\Delta_{2}=1.2$~G,
$B_{2}=78.92$~G. 
Therefore, the window for a phase separation
$a_{12}^{(1d)} < a_{12} < a_{12}^{(mf)}$ 
 translates into the
field range of $93.8$ to $96.3$~G. 
Taking $\omega_{\perp}=2\pi\times 10^{5}$~Hz, we obtain from
Eq.\ (\ref{dilute}) that at the typical particle density $\rho=10^{4}\rm cm^{-1}$ the
phase separation instability will appear already at $B=B_{c}\simeq 94.9$~G,
i.e., before reaching the mean-field threshold value $B_{mf}\simeq 96.3$~G. 
Since recent experiments give hope to the possibility to
control the interspecies scattering length $a_{12}$ within a precision better
than one Bohr radius  \cite{Thalhammer+08}, a detection of this effect must be
within the experimental reach. 

A similar estimate can be done for the $\rm{}^{87}Rb$-$\rm{}^{41}K$ mixture in a
two-dimensional trap. The couplings in that case
\cite{PetrovHolzmannShlyapnikov00} are given by 
$u_{\alpha\beta}=(4\pi\hbar^{3}\omega_{\perp}/m_{\alpha\beta})^{1/2}a_{\alpha\beta}$.
Using a rough estimate $\Lambda_{0}\sim 1/r_{\rm Rb}$, where $a_{\rm Rb}\simeq
248$~pm is the atomic radius of $\rm Rb$, one can obtain from
Eqs.\ (\ref{d2-lc}), (\ref{d2-rhoc}) that at the same transverse confining
frequency  $\omega_{\perp}=2\pi\times 10^{5}$~Hz and a
typical density $\rho=10^{5}cm^{-2}$ the phase separation instability develops
at the critical value of $a_{12}\simeq 54.6$, which with the help of
Eq.\ (\ref{a12}) translates 
into the critical field $B\simeq91.4$~G, well below the mean field value of
$96.3$~G (and this difference can be increased by lowering the density).

\begin{acknowledgments}
 I thank  F. Heidrich-Meisner and U. Schollw\"ock 
for fruitful
discussions. Support by 
 Deutsche For\-schungs\-gemeinschaft (the Heisenberg Program, KO~2335/1-2) is
 gratefully acknowledged.
\end{acknowledgments}

\end{document}